# Characterization of a cylindrical plastic $\beta$-detector with Monte Carlo simulations of optical photons

V. Guadilla[a,*], A. Algora[a,b], J.L. Tain[a], J. Agramunt[a], J. Äystö[c], J.A. Briz[d], A. Cucoanes[d], T. Eronen[c], M. Estienne[d], M. Fallot[d], L.M. Fraile[f], E. Ganioğlu[g], W. Gelletly[a,h], D. Gorelov[c], J. Hakala[c], A. Jokinen[c], D. Jordan[a], A. Kankainen[c], V. Kolhinen[c], J. Koponen[c], M. Lebois[i], T. Martinez[e], M. Monserrate[a], A. Montaner-Pizá[a], I. Moore[c], E. Nácher[j], S.E.A. Orrigo[a], H. Penttilä[c], I. Pohjalainen[c], A. Porta[d], J. Reinikainen[c], M. Reponen[c], S. Rinta-Antila[c], B. Rubio[a], K. Rytkönen[c], T. Shiba[d], V. Sonnenschein[c], E. Valencia[a], V. Vedia[f], A. Voss[c], J.N. Wilson[i], A.-A. Zakari-Issoufou[d]

[a]*Instituto de Física Corpuscular, CSIC-Universidad de Valencia, E-46071, Valencia, Spain*
[b]*Institute of Nuclear Research of the Hungarian Academy of Sciences, Debrecen H-4026, Hungary*
[c]*University of Jyvaskyla, Department of Physics, P.O. Box 35, FI-40014 University of Jyvaskyla, Finland*
[d]*Subatech, CNRS/IN2P3, Nantes, EMN, F-44307, Nantes, France*
[e]*Centro de Investigaciones Energéticas Medioambientales y Tecnológicas, E-28040, Madrid, Spain*
[f]*Universidad Complutense, Grupo de Física Nuclear, CEI Moncloa, E-28040, Madrid, Spain*
[g]*Department of Physics, Istanbul University, 34134, Istanbul, Turkey*
[h]*Department of Physics, University of Surrey, GU2 7XH, Guildford, UK*
[i]*Institut de Physique Nuclèaire d'Orsay, 91406, Orsay, France*
[j]*Instituto de Estructura de la Materia, CSIC, E-28006, Madrid, Spain*

**Abstract**

In this work we report on the Monte Carlo study performed to understand and reproduce experimental measurements of a new plastic $\beta$-detector with cylindrical geometry. Since energy deposition simulations differ from the experimental measurements for such a geometry, we show how the simulation of production and transport of optical photons does allow one to obtain the shapes of the experimental spectra. Moreover, taking into account the computational effort associated with this kind of simulation, we develop a method to convert the simulations of energy deposited into light collected, depending only on the interaction point in the detector. This method represents a useful solution when extensive simulations have to be done, as in the case of the calculation of the response function of the spectrometer in a total absorption $\gamma$-ray spectroscopy analysis.

## 1. Motivation

In $\beta$-decay experiments, $\beta$-detectors are frequently used in coincidence with neutron and/or $\gamma$ detectors in order to clean the measurement by selecting only the events coming from the decays. This method of rejecting the background has been applied mainly with silicon detectors and plastic scintillator detectors. It is important to maximize the $\beta$-detection efficiency in order to maximize the statistics and lower the detection limit, as has been shown in different experimental set-ups with germanium detectors [1, 2] or neutron counters [3]. In the particular case of a Total Absorption $\gamma$-ray Spectroscopy (TAGS) experiment, as well as in experiments with neutron detector arrays, large $4\pi$ detectors are used in order to maximize the efficiency. If a $\beta$-coincidence condition is then required, the statistics is reduced, since the total efficiencies of these kind of detectors -close to 100% for a spectrometer such as the Decay Total Absorption $\gamma$-Ray Spectrometer (DTAS) [4], and around 50% for a neutron detector array such as the BEta deLayEd Neutron (BELEN) counter [5]-, have to be multiplied by the corresponding efficiency of the

*Corresponding author
 Email address: victor.guadilla@ific.uv.es (V. Guadilla)

$\beta$-detector. In the case of TAGS it is also crucial to know accurately the $\beta$-detection efficiency, which depends strongly on the endpoint energy of the $\beta$ branches, affecting the $\beta$-gated spectrometer response. Because of the continuum nature of $\beta$ radiation, the low energy noise discrimination threshold results in a large variation of efficiency over a wide endpoint energy range.

A simple and convenient way to maximize the $\beta$-detection efficiency, minimizing at the same time the $\gamma$ sensitivity, is to build a hollow cylinder of thin plastic scintillation material surrounding the source [1, 2, 3]. Closing one end of the cylinder with scintillation material is a practical way to allow the attachment of a photomultiplier tube (PMT) for light readout, while leaving the other end free for transporting the radioactivity inside the detector, as will be shown in Fig. 1. This geometry resembles the shape of a vase and can have a geometrical efficiency close to 100% for sources at the bottom of the vase. A detector like this was designed and built for experiments in conjunction with DTAS [4] and BELEN [5], aimed at the study of the decay of exotic nuclei. In these experiments the production of the isotopes of interest is low, thus the maximization of the efficiency is a critical requirement.

However, a detector with such a geometry has a different response depending on the interaction point of the $\beta$ particle, as will be shown in this work. This is related to the different light collection efficiency for interactions in the lateral walls of the cylinder and in the bottom, where the PMT is coupled. This effect entails a reduced amount of light collected in the photocathode of the PMT from the lateral walls with respect to the bottom. As a result, the shape of the recorded $\beta$ spectrum is modified, and also makes it difficult to quantify the threshold. This affects dramatically the efficiency curve of the $\beta$-detector, as already mentioned. Therefore the quantification of the efficiency as a function of the endpoint energy cannot be done on the basis of the energy deposited in the detector alone, as we will see, and has to take into account the generation and transport of the light in the detector.

A Monte Carlo (MC) simulation of the light transport can be used to model the response of scintillation detectors. The light transport depends on a number of parameters such as the index of refraction of all optical media, the quality of optical reflectors, the state of surface finish, and light absorption and dispersion in the medium, which are not always known. However it is possible to adjust the parameters empirically to obtain a good reproduction of the measured response [6]. A drawback of the MC simulation of light transport is the long computing time required. This can be an inconvenience when the response has to be computed for a large number of end point energies, as will be explained for the TAGS technique. Therefore we have developed a parametric method that allows the direct conversion of the energy deposited into light collected for this specific geometry.

The cylindrical plastic detector is described in Section 2, and the measurements in Section 3. The MC simulations and the parametric method are discussed in Sections 4 and 5, respectively.

## 2. Geometry and characteristics

The new cylindrical detector can be seen in Fig. 1. It is a vase-shaped cylinder of 35 mm external diameter, 50 mm length and a wall thickness of 3 mm, made of EJ200 plastic scintillator, and manufactured by Scionix [7]. The bottom of the vase is optically glued to a 10 mm length Poly(methyl methacrylate) (PMMA) light guide with two diameters: the front one equal to the plastic detector diameter and the rear one of 39 mm, as can be seen in Fig. 2 bottom. The optical coupling between the light guide and the PMT was made with optical grease. We use a segmented 2×2 multi-anode Hamamatsu PMT R7600U-M4 [8] with an effective area of 18 mm×18 mm, bi-alkali photocathode and 10-stage metal channel dynode structure, operated at an overall voltage of -800 V. The inner walls of the plastic detector were covered by a thin aluminized-mylar reflector in order to improve the light collection.

## 3. Experimental measurements

The four outputs of the PMT were wired in pairs for simplicity. These two signals were integrated in CANBERRA 2005 preamplifiers [9], before being split into two branches for energy and timing reconstruction. In the energy branch the two preamplifier outputs were added and shaped with a CANBERRA 243 amplifier [9], whereas in the timing branch each preamplifier signal was processed independently with an ORTEC 474 Timing Filter Amplifier and an ORTEC 584 Constant Fraction Discriminator [10]. Both timing signals were combined in an ORTEC C04020 Quad 4-input Logic module [10] where a coincidence in a narrow time interval (20 ns) was required to fire the trigger of the data acquisition system. The coincidence requirement allowed the reduction of the noise level and lowered the energy threshold for $\beta$ signals.

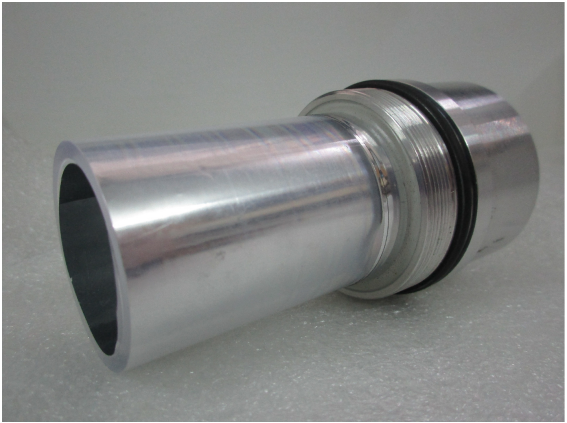

Figure 1: View of the cylindrical plastic $\beta$-detector with the aluminized-mylar reflector inside. The plastic is coupled to a light guide, and mounted in a support of aluminium designed to fit the end of the beam pipe at the IGISOL facility (Jyväskylä).

This new plastic detector was used in the commissioning of the DTAS detector at the upgraded IGISOL IV facility of the University of Jyväskylä, Finland [11], and it was characterized with a a set of calibration sources ($^{22}$Na, $^{60}$Co, $^{24}$Na and $^{137}$Cs). The sources were placed at the bottom of the detector, held by a cylinder of 3M reflector material introduced inside the detector. The detector was later used in the measurement of the $\beta^-$ decay of $^{100}$Tc [12]. In this experiment, the ions were produced by (p,n) reactions on a Mo target situated in the ion source of the IGISOL mass separator. The A=100 separated beam was purified in the double Penning-Trap JYFLTRAP [13] to select $^{100}$Tc, and the nuclei were implanted on the aluminized-mylar reflector that covered the bottom of the detector. Since $^{100}$Tc decays to the stable $^{100}$Ru this did not represent an inconvenience.

## 4. MC simulations

Simulations were carried out with the Geant4 code [14]. The geometry of the detector was defined in great detail, as can be seen in Fig. 2. This is important mainly for the analysis of the TAGS data. The geometry of the PMT was included following the data sheet description [15] and the information provided by Hamamatsu [16]. The photocathode structure and estimated thickness of the dead material were based on Chapter 7 of reference [17]. The internal aluminized-mylar cover is included, as well as the 3M reflector cylindrical support that was used to hold the sources. The 3M reflector composition was estimated based on the details given in [18].

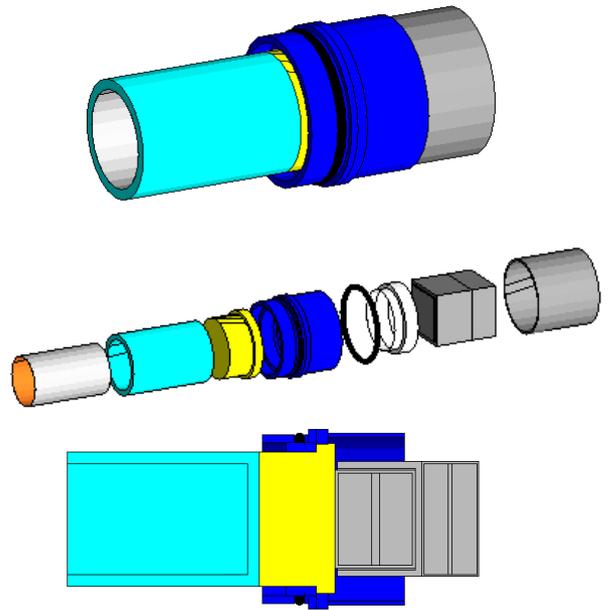

Figure 2: Geometry of the cylindrical $\beta$ plastic detector included in the MC simulations. General view (up), separated components (middle) and lateral cut (bottom) are shown. The following elements are depicted: scintillator material (light blue), light guide (yellow), aluminium support (dark blue), o-ring (black), internal plastic holder (white), PMT plus PMT plastic holder (grey), and aluminized-mylar internal reflector (light grey and orange). (For interpretation of the references to color in this figure caption, the reader is referred to the web version of this paper.)

The task of simulating the light response includes the production in the scintillator material and the transport of the resulting photons until they are absorbed in the photocathode of the PMT. The properties of the scintillator were taken from the EJ200 data sheet [19]: scintillator yield of 10 photons/keV, fast time constant of 0.9 ns and a refraction index of n=1.58. The refraction index of the PMMA and the PMT glass window were set to 1.49 and 1.47 respectively. It is of great importance to define properly the optical interfaces in Geant4 [20, 21], since the result of the simulation is sensitive to the properties of the optical interfaces between the different materials [6]. We defined as dielectric-dielectric all the surfaces where we expected transmission and as dielectric-metal all those where we expected reflection (in our case the interface between the plastic and the aluminized-mylar). All the surfaces were assumed to be polished. The reflectivity of the reflector was set to 1. In order to count the number of photons collected in the photocathode, i.e. photons that suffer an absorption process there, we assigned to the photocathode a constant absorption length of $10^{-9}$ m.

We compare in Fig. 3 the experimental spectrum with the simulation of the energy deposited and of the light collected, both for a calibration source of $^{24}$Na and for the $^{100}$Tc decay. We have used the DECAYGEN event generator [22] to simulate the primary particles of the decays according to the information available ($\gamma$ and $\beta$ intensities, branching ratios, etc). As can be observed, while the simulated energy distribution fails to reproduce the measured spectrum, the simulated light distribution matches the experiment well, specially at low energies, where we are interested in identifying the threshold as will be explained later. In particular, a low energy bump can be distinguished coming from the interaction of the $\beta$ particles in the lateral walls of the detector. Only a small fraction of the light produced there reaches the PMT, thus producing the bump. The higher part of the light distribution above this bump is largely due to the interactions in the bottom of the vase.

We did similar simulations for a second $\beta$-detector with planar geometry, a scintillator disk of 3 mm thickness and 35 mm diameter made of EJ212. This detector was used in another TAGS experiment at IGISOL [23, 24] to measure the $\beta$-s of the decay of several fission fragments in coincidence with the DTAS detector. The $\beta$ spectrum of the decay of $^{140}$Cs measured with this detector is shown in Fig. 4. In this case there are no essential differences between simulations of the light collected and energy deposited, and both reproduce the measurement.

## 5. Energy-Light parametrization

The main problem associated with simulations with optical photons is that they are computationally very demanding. We needed 17 hours of CPU time to perform a $10^5$ events simulation with optical photons in an Intel Core i7-4770 CPU @ 3.40 Hz × 8 with 15.6 GB memory, in contrast to the 50 seconds for a $10^5$ events simulation of the energy deposited. This introduces a disadvantage for the particular case of the TAGS technique, since systematic $\beta$ distribution simulations have to be performed up to the $Q_\beta$ of the decay in small steps (typically 40 keV) to construct the response function of the spectrometer [25]. For this reason a method of avoiding the simulation of optical photons has been developed. This method is based on a relationship found between the energy deposited at different locations in the scintillator material and the light collected in the photocathode. As can be seen in Fig. 5, when we plot light collected versus energy deposited for the $^{100}$Tc simulation, most of the points lie in two well separated regions, depending on whether the energy is deposited in the bottom of the detector or in the lateral walls. These regions show a nearly linear dependence.

This relationship enables us to reproduce the experimental spectra just with the information of the energy deposited and the interaction point. For this we consider two functions, one for the energy deposited in the bottom and the other for the lateral walls. Each of these two functions, in turn, consists of a piecewise function with two quadratic regions:

$$L = \begin{cases} a_1 + b_1 E + c_1 E^2 & , \quad E \leq k \\ a_2 + b_2 E + c_2 E^2 & , \quad E > k \end{cases} \quad (1)$$

where $L$ is the light, $E$ the energy and $k$ the energy value that separates both regions, where continuity is required.

The calibration coefficients obtained for Eq. 1 in the case of the lateral walls and of the bottom of the detector, are reported in Table 1, and they correspond to the red lines drawn in Fig. 5.

In order to apply the conversion procedure it is not enough to use the calibration coefficients in Table 1. We explain, with the help of Fig. 6, the different steps necessary to reproduce the light collection simulation with energy deposited. This figure shows a simulation of $10^4$ events of mono-energetic electrons of 1 MeV interacting with the cylindrical detector. First, we use the calibration coefficients from Table 1 to convert energy into light. As already mentioned, it does not reproduce light simulations, as can be seen in Fig. 6 (a). Second, a Gaussian spread, empirically found to be proportional to $E^{3/4}$, is introduced, as shown in Fig. 6 (b), in order to reproduce the width of the peaks produced by the interaction of the mono-energetic electrons. Finally, for 10% of the events interacting with the bottom of the detector we introduce a random loss of light collected. This corresponds to events in between both regions in Fig. 5. For this, we change the slope of the calibration with a random linear interpolation between the slope of the bottom and the slope of the lateral walls. It improves the comparison with light simulations, as can be seen in Fig. 6 (c). The result of following these steps, combining the calibration in Table 1 with the corrections explained here, is good agreement between light simulations, energy simulations converted to light, and experimental measurements, as can be observed in Fig. 3 for both the $^{24}$Na calibration source and the $^{100}$Tc decay.

In order to apply this procedure to calculate the $\beta$ responses for a TAGS analysis, we have to convert the threshold in signal amplitude, or equivalently in light collected, into a threshold in energy, according to the calibration in Table 1. A threshold value in light of 11 a.u. has been identified by comparison with experimental measurements in Fig. 3. Due to the effect of the different light collection in the two regions of the detector, this threshold corresponds to 29 keV in the bottom, whereas for the lateral walls of the detector it is 96 keV. To verify this equivalence, we simulated four end-points of the $^{100}$Tc decay, namely 3203 keV, 2072 keV, 1151 keV and 543 keV, and we calculated the efficiency above this threshold value for the light simulation and for the energy converted into light. Both efficiencies are in very good agreement, as shown in Fig. 7, with relative differences of $\sim 0.1\%$.

## 6. Conclusions

MC simulations with optical photons have been shown to reproduce the shape of the experimental $\beta$ spectra measured with a plastic scintillation detector with a vase-shaped geometry, when energy deposited simulations turned out to be insufficient. Furthermore, a method to directly convert the energy deposited into the equivalent amount of light has been developed, and successfully applied to some experimental cases. This opens the possibility of avoiding the very time-consuming simulations with optical photons for this type of detector. In the particular case of a TAGS analysis, where we are interested in doing extensive simulations of $\beta$ particles in steps of 40 keV up to the $Q_\beta$, this option provides an affordable way to calculate the response function of the spectrometer. In fact, this method has been applied to the TAGS analysis of the measurement of the $\beta^-$ decay of $^{100}$Tc, that was mentioned in this work and will be published in the near future [26].

## 7. Acknowledgements

This work has been supported by the Spanish Ministerio de Economía y Competitividad under grants FPA2011-24553, AIC-A-2011-0696, FPA2014-52823-C2-1-P and the program Severo Ochoa (SEV-2014-0398), by the European Commission under the FP7/EURATOM contract 605203, and by the Spanish Ministerio de Educación under the FPU12/01527 grant. Helpful discussions with P. Schotanus (Scionix) and the development of the detector by Scionix are also acknowledged.

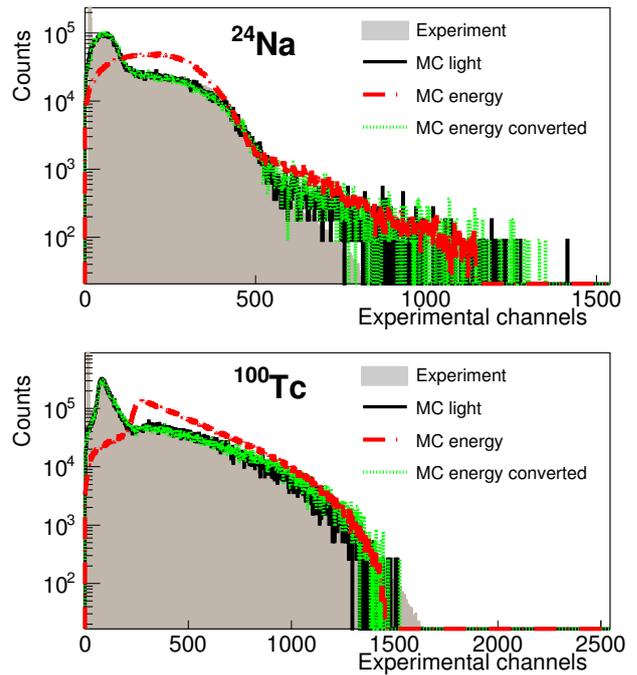

Figure 3: Comparison between the experimental and the MC spectra for $^{24}$Na (up) and $^{100}$Tc (bottom). Experiment (grey filled) is compared with simulations of energy deposited (dashed-dotted red), light collected (solid black), and energy deposited converted into light with the procedure explained in the text (dotted green). Note that energy simulations (dashed-dotted red) are performed with $10^6$ events, while the rest correspond to $10^5$. (For interpretation of the references to color in this figure caption, the reader is referred to the web version of this paper.)

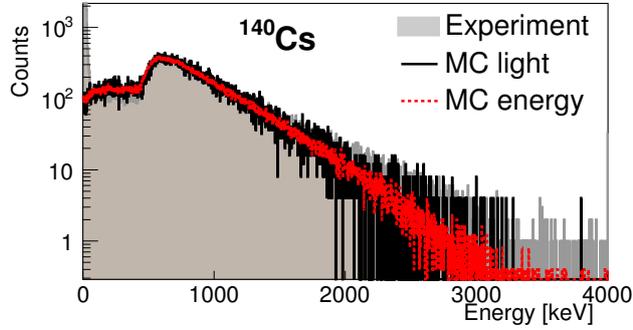

Figure 4: Comparison between the experimental and the MC spectra for a $^{140}$Cs measurement performed with a plastic scintillator disk. Experiment (grey filled) is compared with simulations of energy deposited (dotted red) and light collected (solid black). Note that energy simulations (red) are performed with $10^6$ events, while the simulations of light (blue) correspond to $10^5$. (For interpretation of the references to color in this figure caption, the reader is referred to the web version of this paper.)

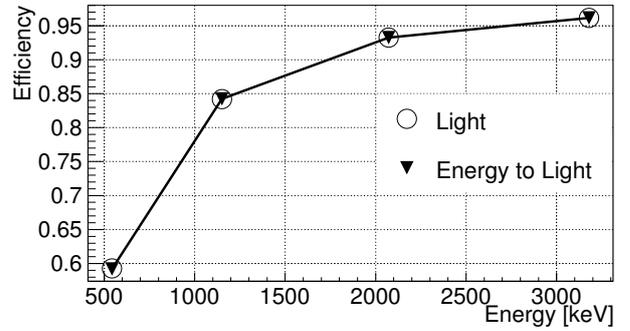

Figure 7: Efficiency of the cylindrical plastic detector for different $\beta^-$ end-points. The results for light simulations (circles) are compared with the calculation using our procedure for converting energy deposited into light collected (triangles).

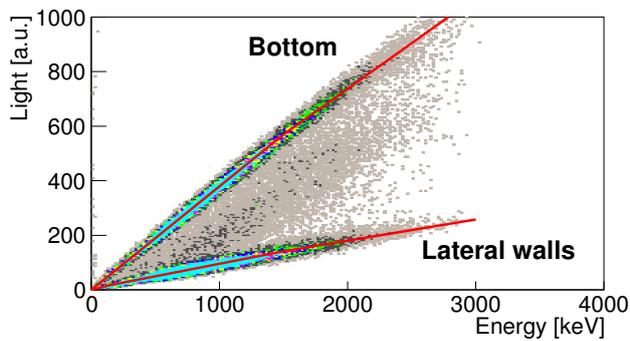

Figure 5: Simulation of the light collected vs. energy deposited for the cylindrical plastic detector in the $^{100}$Tc measurement. Two different regions are distinguished and the calibration curves with the parameters from Table 1 are represented in red. The events in between both regions correspond to a 10% of those coming from the bottom, where less light than expected is collected.

| Part | $a_1$ [a.u] | $b_1$ [a.u. keV$^{-1}$] | $c_1$ [a.u. keV$^{-2}$] | $k$ [keV] | $a_2$ [a.u.] | $b_2$ [a.u. keV$^{-1}$] | $c_2$ [a.u. keV$^{-2}$] |
|---|---|---|---|---|---|---|---|
| Lateral | 0.0 | 0.1 | 0.0001 | 150 | 2.3625 | 0.1 | -0.000005 |
| Bottom | 0.0 | 0.38 | 0.0 | 1500 | 22.5 | 0.38 | -0.00001 |

Table 1: Energy-light calibration parameters for the lateral walls and the bottom of the cylinder following the Eq. 1. The parameter $k$ separates two different energy regions.

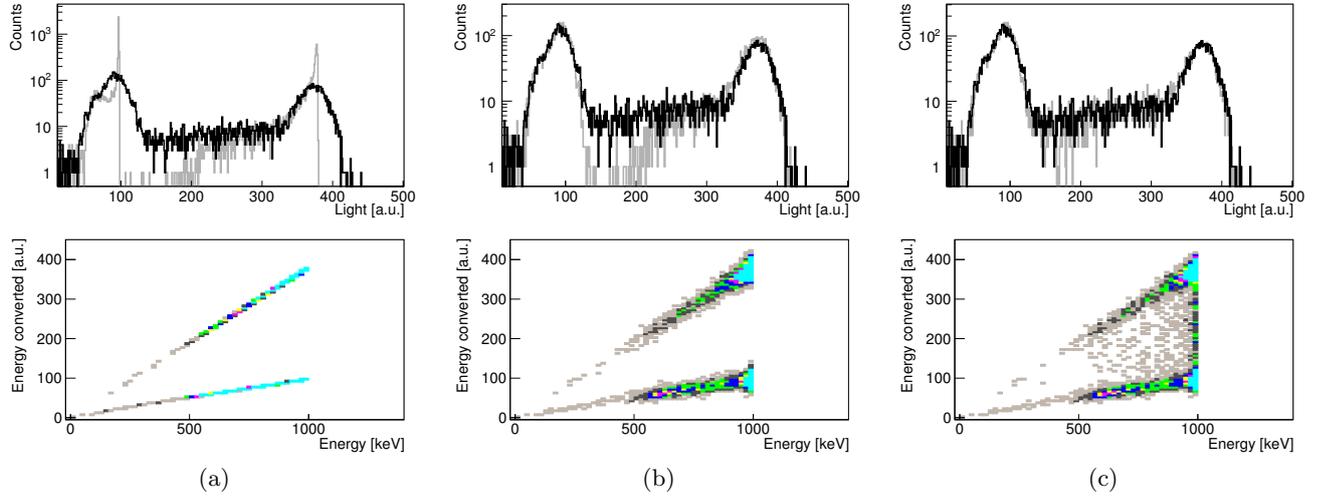

(a)          (b)          (c)

Figure 6: Figures (a), (b) and (c) represent different steps necessary to reproduce the simulated light collection with the results of the simulated energy deposition (for more details see the text). In the upper panel the simulation of optical photons (black) is compared with the energy converted into light (grey) for mono-energetic electrons of 1 MeV interacting with the detector. The lower panel shows the corresponding relationship between energy converted into light and energy deposited for this case.